\documentclass[aps,prd,floatfix,superscriptaddress,preprintnumbers]{revtex4}
\usepackage{amssymb}
\usepackage{amsmath}
\usepackage{amsfonts}
\usepackage{epsfig}             
\usepackage{graphicx}
\usepackage{tabularx}
\usepackage{color}
\newcommand{\seq}{\begin{subequations}}
\newcommand{\sen}{\end{subequations}}
\newcommand{\eq}{\begin{eqnarray}}
\newcommand{\en}{\end{eqnarray}}
\newcommand{\bfq}{{\bf q}_{\perp}}
\newcommand{\bfk}{{\bf k}_{\perp}}

\begin{document}

\title{Scaling of PDFs, TMDs, and GPDs in soft-wall AdS/QCD} 

\author{Valery E. Lyubovitskij} 
\affiliation{Institut f\"ur Theoretische Physik,
Universit\"at T\"ubingen, 
Kepler Center for Astro and Particle Physics,  
Auf der Morgenstelle 14, D-72076 T\"ubingen, Germany}
\affiliation{Departamento de F\'\i sica y Centro Cient\'\i fico
Tecnol\'ogico de Valpara\'\i so-CCTVal, Universidad T\'ecnica
Federico Santa Mar\'\i a, Casilla 110-V, Valpara\'\i so, Chile}
\author{Ivan Schmidt}
\affiliation{Departamento de F\'\i sica y Centro Cient\'\i fico
Tecnol\'ogico de Valpara\'\i so-CCTVal, Universidad T\'ecnica
Federico Santa Mar\'\i a, Casilla 110-V, Valpara\'\i so, Chile}

\date{\today}

\begin{abstract}

We explicitly demonstrate how to correctly define the 
hadronic parton distributions (PDFs, TMDs, and GPDs) in the
soft-wall AdS/QCD approach, based on the use of a quadratic 
dilaton field, providing confinement and 
breaking of conformal and chiral symmetries. 
The power behavior of parton distributions 
at large values of the light-cone variable is 
consistent with quark counting rules and Drell-Yan-West duality. 
All parton distributions are defined in terms of profile functions,  
which depend on the light-cone coordinate 
and are fixed from PDFs and electromagnetic form factors. 

\end{abstract}

\maketitle

\section{Introduction}

During the last decade the soft-wall AdS/QCD~\cite{Karch:2006pv}-\cite{Andreev:2006vy} 
formalism achieved significant progress in the description of hadron structure: 
mass spectrum, parton distributions, form factors, etc. 
(for overview, see, e.g., Ref.~\cite{Brodsky:2014yha}),
based on an effective action constructed 
with the use of a quadratic dilaton field providing confinement and 
breaking of both conformal and chiral symmetry. This dilaton field $\varphi(z)$ has 
quadratic dependence on the holographic variable $z$, and is multiplied with 
the dilaton scale parameter $\kappa$ (of order of a few hundreds of MeV): 
$\varphi(z) = \exp(-\kappa^2 z^2)$. 
Chiral symmetry breaking can be achieved  in the soft-wall AdS/QCD model by using 
a modified dilaton profile and a quartic term in the bulk scalar potential.  
Such a modification allows to separate the dependence on spontaneous and 
explicit chiral symmetry breaking. 
One should stress that in Ref.~\cite{Gherghetta:2009ac} the issue of 
chiral symmetry breaking was further studied and understood. 
In particular, in~\cite{Jarvinen:2011qe} a new class of holographic 
models has been derived in the so-called
Veneziano limit, in which both the number of flavors
and of colors are large $N_f, N_c \to \infty$ and their ratio
$N_f/N_c$ is fixed. In this approach, chiral symmetry
breaking is ruled out by the running of the anomalous dimension of 
the chiral condensate and 
the Breitenlohner-Freedman (BF) bound violation.

Soft-wall AdS/QCD is mainly phenomenological approach in the class o
f holographic 
approaches and needs in further justification to be more consistent with QCD. 
However, there are a few advantages of this approach, which make it usefull 
in study of properties of hadrons and exotic states. 
One of its advantages is that it explicitly 
reproduces the power scaling of hadronic form factors at large 
$Q^2$~\cite{Brodsky:2014yha}-\cite{Gutsche:2016lrz}. 
In particular, soft-wall AdS/QCD is consistent with the Drell-Yan-West (DYW) 
relation~\cite{Drell:1969km} 
between the large-$Q^2$ behavior of nucleon electromagnetic form factors and 
the large-$x$ behavior of the structure functions (see also 
Ref.~\cite{Bloom:1970xb} for the extension to inelastic scattering) and 
quark counting rules~\cite{Brodsky:1973kr}. Based on the findings 
in Refs.~\cite{Drell:1969km,Bloom:1970xb,Brodsky:1973kr} one can, e.g.,  
relate the behavior of the quark distribution function (PDF) in nucleon 
$q_v(x) \sim (1-x)^p$ at $x \to 1$ with the scaling of the proton Dirac 
form factor $F_1^p(Q^2) \sim 1/(Q^2)^{(p+1)/2}$ at large $Q^2$, where 
the parameter $p$ is related to the number of constituents in the proton 
(or twist $\tau$) as 
$p = 2 \tau - 3$~\cite{Drell:1969km,Blankenbecler:1974tm}. 
At large $x$ and finite $Q^2$ 
there are also model-independent predictions of perturbative QCD (pQCD) for 
the generalized parton distributions (GPDs)~\cite{Yuan:2003fs} --- 
pion ${\cal H}_q^\pi(x,Q^2)$ and nucleon ${\cal H}_q^N(x,Q^2)$, 
${\cal E}_q^N(x,Q^2)$: 
\eq
\label{Hpi_N}
{\cal H}_q^\pi(x,Q^2) \sim (1-x)^2\,, \quad 
{\cal H}_q^N(x,Q^2)   \sim (1-x)^3\,, \quad {\cal E}_q^N(x,Q^2) 
\sim (1-x)^5 \,.
\en 
Note that the prediction of pQCD for the pion PDF $q_\pi(x) \sim (1-x)^2$ 
at large $x$ 
(it trivially follows from the prediction for GPDs~\cite{Yuan:2003fs}) 
was supported by the updated analysis~\cite{Aicher:2010cb} 
of the E615 data~\cite{Conway:1989fs} on the cross section 
of the Drell-Yan (DY) process $\pi^- N \to \mu^+ \mu^- X$,  
including next-to-leading logarithmic (NLL) threshold resummation effects: 
$q_\pi(x) \sim (1-x)^{2.03}$ at the initial scale 
$\mu_0 = 0.63$ GeV~\cite{Aicher:2010cb}.    

Parton distributions in hadrons play an important role in the QCD description 
of hadrons and in their interactions in terms of quarks and gluons. In particular, 
QCD factorization allows to separate effects
of strong interactions at small distances (perturbative dynamics of quark and gluons)
from long-distance (or small momenta) effects - nonperturbative part. This last part is
parametrized by parton distribution functions, which are
universal functions for each hadron
and independent of the specific process.
In this vein one can represent observable quantities, such as cross sections,
as having both perturbative and nonperturbative pieces. The perturbative 
part of the cross section is defined by those subprocesses which come from the 
hard interactions of quarks, gluons
and electroweak particles and which can then be calculated perturbatibely using the Standard Model. 
The nonperturbative part is encoded in parton distributions, 
which cannot be directly calculated in QCD, and therefore different 
theoretical approaches (world data analysis, lattice and light-front QCD, 
quark and potential models, etc.) 
have been applied to extract or predict the PDFs, TMDs, and GPDs 
(for a recent overview see, e.g. 
Ref.~\cite{Lin:2020rut}). 

The importance of the scaling laws and their role 
in the description of nucleon structure has been stressed and studied 
in detail in 
the literature. Moreover, they are important for the proper construction of 
light-front (LF) QCD approaches~\cite{Brodsky:2014yha,Brodsky:2007hb},
\cite{Vega:2009zb}-\cite{Brodsky:2020ajy}, motivated by soft-wall AdS/QCD 
and developed in the past decade. The main advantage of these 
LF QCD approaches was in the construction of effective 
wave functions for mesons~\cite{Brodsky:2014yha,Brodsky:2007hb,Vega:2009zb,%
Branz:2010ub,Brodsky:2011xx,Gutsche:2014zua,deTeramond:2018ecg,%
Brodsky:2020ajy,Vega:2020ctz}, 
baryons~\cite{Lyubovitskij:2013ski,Gutsche:2013zia,Gutsche:2014yea,Gutsche:2016gcd,%
deTeramond:2018ecg,Brodsky:2020ajy}, and for hadrons with arbitrary number of partons 
(arbitrary twist)~\cite{Lyubovitskij:2013ski,Gutsche:2013zia,Gutsche:2014yea,%
deTeramond:2018ecg,Brodsky:2020ajy}, which were further used in the calculation 
of fundamental properties of hadrons - parton distributions and form factors. 
While form factors and parton distributions in LF QCD were consistent with 
quark counting rules at large $Q^2$ and large $x \to 1$ (light-cone variable), 
there was the problem of achieving full consistency in soft-wall AdS/QCD. 
As we stressed before, hadronic form factors in soft-wall AdS/QCD obey 
the power scaling $1/Q^{2 (\tau-1)}$ at large $Q^2$ and for arbitrary twist $\tau$ 
of a hadron. On the other hand, parton distributions (like PDF and GPDs) 
calculated in soft-wall AdS/QCD 
(see, e.g., Refs.~\cite{Brodsky:2007hb,Vega:2010ns,Gutsche:2011vb}) 
have different scaling at large $x$. In particular, the pion PDF scaled 
as $(1-x)^0$~\cite{Brodsky:2007hb,Gutsche:2011vb}, the nucleon charged and 
magnetization PDFs/GPDs are scaled as $(1-x)$ and $(1-x)^2$, 
respectively~\cite{Vega:2010ns}. Such behavior of PDFs was obtained starting from 
the effective actions for mesons and baryons with total 
angular spin $J$ ~\cite{Brodsky:2014yha}-\cite{Gutsche:2016lrz} (see 
in the next section discussion of these actions and derivation of the PDFs/GPDs from 
them).  

In Refs.~\cite{Brodsky:2007hb,Vega:2010ns,Gutsche:2011vb} 
the integral representation for the hadronic 
form factor with twist $\tau$ has been derived, 
which can also be written in closed form 
as the beta function $B(\alpha,\beta)$
\eq 
F_\tau(Q^2) = \int\limits_0^1 dy \, (\tau - 1) \, (1-y)^{\tau-2} \, y^a 
= (\tau-1) \, B(\tau-1,a+1) 
\en 
Using identification of the $y$ variable with the light-cone momentum fraction $x$ 
both PDFs $q_\tau(x)$ and
GPDs ${\cal H}_\tau(x,Q^2)$ have 
been extracted~\cite{Brodsky:2007hb,Vega:2010ns,Gutsche:2013zia}: 
\eq 
q_\tau(x) = (\tau - 1) \, (1-x)^{\tau-2}\,, \quad {\cal H}_\tau(x,Q^2) = q_\tau(x) \, x^a \,. 
\en 
Such $x$ dependence of PDF and GPD contradicts 
model-independent results: the DY inclusive counting rule 
for $q_\tau(x)$ at $x \to 1$~\cite{Drell:1969km,Blankenbecler:1974tm,Yuan:2003fs} 
and the prediction of pQCD for GPDs  --- pion ${\cal H}_q^\pi(x,Q^2)$ 
and nucleon ${\cal H}_q^N(x,Q^2)$, ${\cal E}_q^N(x,Q^2)$ at large $x$ and 
finite $Q^2$~\cite{Yuan:2003fs}. 

It was first noticed in Ref.~\cite{Lyubovitskij:2013ski} that the interpretation 
of the variable $y$ 
in the integral representation~(\ref{VInt}) as light-cone variable is not truly correct 
and that one can think about a {\it generalized light-cone variable $y(x)$} depending on $x$. 
Then the power behavior of hadronic PDFs and GPDs ar large $x$ is consistent 
with model-independent 
results of Refs.~\cite{Drell:1969km,Blankenbecler:1974tm,Yuan:2003fs} can be obtained, 
provided that an appropriate choice of the $x$ dependence of the function $y(x)$ is made. 
In particular, the simplest choice the function $y(x)$ was found as 
\eq\label{ytau} 
y_N(x) = \exp\Big[-\log(1/x) (1-x)^{2/(N-1)}\Big]
\en 
leading to the correct large-$x$ scaling of PDFs and GPDs in mesons  
\eq 
q_\tau^M(x) \sim {\cal H}_\tau^M(x,Q^2) \sim (1-x)^{2 \tau - 2}
\en 
at $N=2 \tau - 2$ and in baryons 
\eq 
q_\tau^B(x) \sim {\cal H}_\tau^B(x,Q^2) \sim (1-x)^{2 \tau - 3}
\en 
at $N=2 \tau - 3$. The function $y_\tau(x)$ obeys the following boundary 
conditions $y_\tau(0) = 0$ and $y_\tau(1) = 1$.  
Notice that a similar idea was recently considered in the framework of 
light-front holographic QCD (LFHQCD)~\cite{deTeramond:2018ecg,Brodsky:2020ajy} (see also 
Ref.~\cite{Chang:2020kjj} for an extension of Ref.~\cite{deTeramond:2018ecg}). 
In particular, a function [named as $w(x)$] was introduced in the integral 
representation of the form factor~\cite{deTeramond:2018ecg,Brodsky:2020ajy}: 
\eq 
F_\tau(Q^2) = \frac{1}{N_\tau} \, \int\limits_0^1 dx \, w^\prime(x) \, 
[w(x)]^{Q^2/4\lambda-1/2} \, [1-w(x)]^{\tau-2}
\en 
Obviously, both mathematical extensions considered in Refs.~\cite{Lyubovitskij:2013ski}   
and~\cite{deTeramond:2018ecg,Brodsky:2020ajy} are equivalent. The only difference is that 
in Refs.~\cite{deTeramond:2018ecg,Brodsky:2020ajy} an extra power $-1/2$ was included 
in the $[w(x)]^{Q^2/4\lambda-1/2}$, while in the soft-wall 
model~\cite{Brodsky:2007hb,Vega:2010ns,Gutsche:2011vb} the factor 
is $[w(x)]^{Q^2/4\lambda}$. In other words, the soft-wall 
model~\cite{Brodsky:2007hb,Vega:2010ns,Gutsche:2011vb} 
and LFHQCD~\cite{Brodsky:2007hb,Vega:2010ns,Gutsche:2011vb} deal with slightly 
different analytical expressions for the hadronic form factors: 
$F_\tau(Q^2) \sim B(\tau - 1,1+Q^2/4\lambda)$ in soft-wall AdS/QCD 
and $F_\tau(Q^2) \sim B(\tau - 1,1/2+Q^2/4\lambda)$ in LFHQCD.  

The main objective of this paper is to continue the discussion of ideas started 
in Ref.~\cite{Lyubovitskij:2013ski,deTeramond:2018ecg,Brodsky:2020ajy} 
and propose a more simple derivation of PDFs, TMDs, and GPDs of hadrons 
with arbitrary twist in the context of soft-wall AdS/QCD models. 
In particular, we explicitly demonstrate how to correctly define the 
hadronic parton distributions (PDFs, TMDs, and GPDs) in the
soft-wall AdS/QCD approach, based on the use of a quadratic 
dilaton field providing confinement and 
breaking of conformal and chiral symmetry. 
The obtained power behavior of parton distributions at large values 
of light-cone variable $x$ are then consistent 
with quark counting rules and DYW duality. 
All parton distributions are defined in terms of profile functions depending on 
the light-cone coordinate and are fixed from PDFs and electromagnetic form factors. 

The paper is organized as follows.
In Sec.~II we present overview of our approach and 
consider derivation of PDFs. 
TMDs will be derived in Sec.~III. 
In Sec.~IV we discuss derivation of GPDs. 
Finally, Sec.~V contains our summary and conclusions. 

\section{Parton Distribution Functions (PDFs)}

\subsection{General consideration}

We start with a brief overview of the effective action for 
AdS fields (bosons and fermions) dual to mesons and baryons, 
derived in our previous papers (for more details 
see, e.g., Refs.~\cite{Vega:2010ns,Gutsche:2011vb}). 
in the next section). We consider the propagation of boson 
$\Phi_{M_1 \cdots M_J}(x,z)$ and fermion $\Psi_{M_1 \cdots M_{J-1/2}}(x,z)$ 
fields with spin $J$ 
(dual to mesons and baryons, respectively) in a five-dimensional 
AdS space. The AdS metric is specified by 
\eq
ds^2 =
g_{MN} dx^M dx^N = \eta_{ab} \, e^{2A(z)} \, dx^a dx^b = e^{2A(z)}
\, (\eta_{\mu\nu} dx^\mu dx^\nu - dz^2)\,, \hspace*{1cm}
\eta_{\mu\nu} = {\rm diag}(1, -1, \ldots, -1) \,,
\en
where $M$ and
$N = 0, 1, \cdots , d$ are the space-time (base manifold) indices,
$a=(\mu,z)$ and $b=(\nu,z)$ are the local Lorentz (tangent) indices,
$g_{MN}$ and  $\eta_{ab}$ are curved and flat metric tensors, which
are related by the vielbein $\epsilon_M^a(z)= e^{A(z)} \,                     
\delta_M^a$ as $g_{MN} =\epsilon_M^a \epsilon_N^b \eta_{ab}$. Here
$z$ is the holographic coordinate, $R$ is the AdS radius, and $g =             
|{\rm det} g_{MN}| = e^{10 A(z)}$. We restrict
ourselves to a conformal-invariant metric with $A(z) = \log(R/z)$, 
where $R$ is the AdS radius. 
The boson action is written as: 
\eq
S_B &=& \frac{(-)^J}{2}
\int d^d x dz \sqrt{g} \, e^{-\varphi(z)}
\biggl[ g^{MN} g^{M_1N_1} \cdots g^{M_JN_J} \,
\partial_M\Phi_{M_1 \cdots M_J}(x,z) \,
\partial_N\Phi_{N_1 \cdots N_J}(x,z) \nonumber\\
&-& (\mu_J^2 + V_J(z)) \, g^{M_1N_1} \cdots g^{M_JN_J}
\Phi_{M_1 \cdots M_J}(x,z) \, \Phi_{N_1 \cdots N_J}(x,z) \biggr]
\en 
where bosonic spin-$J$ field $\Phi_{M_1 \cdots M_J}(x,z)$ is 
described 
by a symmetric, traceless tensor, satisfying the conditions
\eq
\partial^{M_1}  \Phi_{M_1M_2 \cdots M_J} = 0\,, \quad\quad
g^{M_1M_2}  \Phi_{M_1M_2 \cdots M_J} = 0\,, 
\en
Here $V_J(z) = e^{-2A(z)} U_J(z)$, where $U_J(z)$ 
is the effective dilaton potential
\eq
U_J(z) = \frac{1}{2} \varphi^{\prime\prime}(z)
\,+\, (d-1-2J) \, \varphi^{\prime}(z) A^{\prime}(z) 
\en 
and 
\eq
\mu_J^2 R^2 = (\Delta - J) (\Delta + J - 4) 
\en 
is the bulk mass. 
The quadratic dilaton field $\varphi(z)$ is specified as 
$\varphi(z) = \kappa^2 z^2$, where $\kappa$ 
is the dimensional parameter. The dimension of the boson 
AdS fields $\Delta$ is identified with twist $\tau$ as 
$\Delta = \tau = N + L$, where $N$ is the number of partons 
and $L$ is the orbital angular momentum. 

Restricting to the axial gauge $\Phi_{\cdots z \cdots}(x,z)=0$ 
and performing the Kaluza-Klein expansion 
\eq\label{KK_coord_PhiJ}
\Phi^{\mu_1 \cdots \mu_J}(x,z)
= \sum\limits_n \ \Phi^{\mu_1 \cdots \mu_J}_n(x) \ \Phi_{n}(z)
\en
one can derive the equation of motion (EOM) for the profile 
function $\phi_{n\tau}(z) =  e^{3A(z)/2} \, \Phi_n(z)$:  
\eq\label{Eq1J}
\Big[ - \frac{d^2}{dz^2} + \frac{4 (\tau-2)^2 - 1}{4z^2} + U_J(z)
\Big] \phi_{n\tau}(z) = M^2_{n\tau\!J} \phi_{n\tau}(z)
\en 
with analytical solutions for the bulk profile 
\eq
\phi_{n\tau}(z) &=& \sqrt{\frac{2}{\Gamma(\tau-1)}} \, \kappa^{\tau-1} \, 
z^{\tau - 3/2} \, e^{-\kappa^2 z^2/2} \, L_n^{\tau-2}(\kappa^2 z^2) 
\en 
and mass spectrum 
\eq\label{mass2_J}
M^2_{n\tau\!J} = 4 \kappa^2 \Big( n + \frac{\tau + J}{2} - 1 \Big) \,.
\en 
Here $L_n^{m}(x)$ are the  
generalized Laguerre polynomials. 

In the case of AdS fermion fields $\Psi_{K_1 \cdots K_{J-1/2}}(x,z)$ 
with spin $J$, the action reads~\cite{Gutsche:2011vb}: 
\eq
\hspace*{-.5cm}
S_{F} &=&  \int d^dx dz \, \sqrt{g} \, e^{-\varphi(z)} \,
g^{K_1N_1} \cdots g^{K_{J-1/2}N_{J-1/2}} \,
\biggl[ \frac{i}{2} \bar\Psi_{K_1 \cdots K_{J-1/2}}(x,z)
\epsilon_a^M \Gamma^a {\cal D}_M \Psi_{N_1 \cdots N_{J-1/2}}(x,z)
\nonumber\\
&-& \frac{i}{2}
({\cal D}_M\Psi_{K_1 \cdots K_{J-1/2}}(x,z))^\dagger
\Gamma^0 \epsilon_a^M \Gamma^a \Psi_{N_1 \cdots N_{J-1/2}}(x,z)
- \bar\Psi_{K_1 \cdots K_{J-1/2}}(x,z) \Big(\mu + V_F(z)\Big)
\Psi_{N_1 \cdots N_{J-1/2}}(x,z) \biggr] \,,
\en
where $V_F(z) = \varphi(z)/R$ is the dilaton potential, 
${\cal D}_M$ is the covariant derivative acting on the spin-tensor
field $\Psi^\pm_{N_1 \cdots N_{J-1/2}}$ as: 
\eq
{\cal D}_M \Psi_{N_1 \cdots N_{J-1/2}} =
\partial_M \Psi_{N_1 \cdots N_{J-1/2}}
- \frac{1}{8} \omega_M^{ab} [\Gamma_a, \Gamma_b]
\Psi_{N_1 \cdots N_{J-1/2}} \,,
\en
where $\omega_M^{ab} = A^\prime(z) \,  
(\delta^a_z \delta^b_M - \delta^b_z \delta^a_M)$ 
is the spin connection term, and 
$\Gamma^a=(\gamma^\mu, - i\gamma^5)$ are the Dirac matrices.

After expanding the fermion field in left- and right-chirality
components $\Psi^{L/R} = (1 \mp \gamma^5)/2 \, \Psi$ and 
a KK expansion for the $\Psi^{L/R}(x,z)$ fields 
$\Psi^{L/R}(x,z) = \sum\limits_n
\ \Psi^{L/R}_n(x) \ F_n^{L/R}(z)$,
one can obtain decoupled Schr\"odinger EOMs for the fermion 
bulk profiles $f_n^{L/R}(z) = e^{2 A(z)} \, F_n^{L/R}(z)$: 
\eq
\biggl[ -\partial_z^2
+ \kappa^4 z^2 + 2 \kappa^2 \Big(m \mp \frac{1}{2} \Big)
+ \frac{m (m \pm 1)}{z^2} \biggr] f_{n\tau}^{L/R}(z) 
= M_{n\tau}^2 \, f_{n\tau}^{L/R}(z) \,, 
\en
where $m = \tau - 3/2 = L + 3/2$ and 
\eq\label{fLR_tau}
f_{n\tau}^{L}(z) &=& \sqrt{\frac{2\Gamma(n+1)}{\Gamma(n+\tau)}} \ \kappa^{\tau}
\ z^{\tau-1/2} \ e^{-\kappa^2 z^2/2} \ L_n^{\tau-1}(\kappa^2z^2) \,, \\
f_{n\tau}^{R}(z) &=& \sqrt{\frac{2\Gamma(n+1)}{\Gamma(n+\tau-1)}} \ \kappa^{\tau-1}
\ z^{\tau-3/2} \ e^{-\kappa^2 z^2/2} \ L_n^{\tau-2}(\kappa^2z^2)
\en
and
\eq
M_{n\tau}^2 = 4 \kappa^2 \Big( n + \tau - 1 \Big) 
            = 4 \kappa^2 \Big( n + L + 2 \Big) \,.  
\en
In order to study electromagnetic properties of hadrons we 
need to calculate the vector bulk-to-boundary propagator $V(q,z)$, dual to 
the $q^2$-dependent electromagnetic current: 
\eq
\partial_z \biggl( \frac{e^{-\varphi(z)}}{z} \,
\partial_z V(-q^2,z)\biggr) + q^2 \frac{e^{-\varphi(z)}}{z} \,
V(-q^2,z) = 0 \,.
\en
The latter equation is solved analytically in terms of 
the gamma $\Gamma(n)$ and Tricomi $U(a,b,z)$ functions: 
\eq
\label{VInt_q}
V(Q^2,z) = \Gamma(1 + a) \, U(a,0,\kappa^2 z^2) \,, 
\en
where $Q^2 = - q^2$ and $a = Q^2/(4 \kappa^2)$. 
It is convenient to use the integral
representation for $V(Q,z)$~\cite{Grigoryan:2007my}
\eq
\label{VInt}
V(Q^2,z) = \kappa^2 z^2 \int\limits_0^1 \frac{dy}{(1-y)^2}
\, y^{a} \,
e^{- \kappa^2 z^2 \frac{y}{1-y} }\,. 
\en
The expression for the hadron form factors in the soft-wall AdS/QCD 
is given by 
\eq
F_{n\tau}(Q^2) = \int\limits_0^\infty dz \, \phi_{n\tau}^2(z) \, V(Q^2,z) \,, 
\en 
where the integrand contains the square of the holographic wave function 
in fifth dimension $z$ (dual to hadron wave function), multiplied with the 
vector bulk-to-boundary propagator $V(Q^2,z)$. 

Now we in a position to start the derivation of PDFs in soft-wall AdS/QCD. 
In the following, for simplicity, we restrict to a consideration 
of ground states of hadrons with $n=0$. 
In Ref.~\cite{Lyubovitskij:2013ski} 
and~\cite{deTeramond:2018ecg,Brodsky:2020ajy}, this quantity has been derived 
using an integral representation for the hadronic form factor (see discussion 
in the Introduction). The easiest way 
is to start with the hadronic wave function normalization condition, which
depends on the holographic variable $z$: 
\eq\label{norm_phi}
1 = \int\limits_0^1 dz \, \phi_\tau^2(z) 
\en 
where $\phi_\tau(z)$ is the AdS bulk profile function (for simplicity 
we restrict here to the bosonic case and extension on fermion 
case is straightforward). 
Next we use the integral representation for unity 
\eq\label{int_unity} 
1 = - e^{\kappa^2 z^2} \, 
\int\limits_0^1 d\biggl[f_\tau(x) \, e^{-\kappa^2 z^2/(1-x)^2} \biggr]
= e^{\kappa^2 z^2} \,   
\int\limits_0^1 dx \, \biggl[ 
\frac{2f_\tau(x) \, \kappa^2 z^2}{(1-x)^3} \, - f'_\tau(x) \biggr] 
\, e^{-\kappa^2 z^2/(1-x)^2} 
\en  
and insert it into Eq.~(\ref{norm_phi}). Here $x$ is the 
light-cone coordinate and $f_\tau(x)$ is the profile function 
with boundary condition $f_\tau(0) = 1$, 
which is specific for a particular hadron and fixed from its PDF. 
The functions $f_\tau(x)$ and $y_\tau(x)$ [see Eq.~(\ref{ytau})] 
are related as: 
\eq 
\Big(1-y_\tau(x)\Big)^{\tau-1} = f_\tau(x) \, (1-x)^{2 (\tau-1)}
\en 
or 
\eq\label{yf_relation} 
y_\tau(x) = 1 - \Big[f_\tau(x)\Big]^{\frac{1}{\tau-1}} \, (1-x)^2 \,. 
\en 
We remind that at $x=0$ the functions $y_\tau(x)$ and $f_\tau(x)$  
obey the boundary conditions $y_\tau(0) = 0$ and $f_\tau(0) = 1$. 
At $x=1$ function $f_\tau$ is finite and its value 
depends on the specific choice of twist $\tau$ (see below), 
while $y_\tau(1) = 1$ is independent on twist.

After integration over the variable $z$ we get 
\eq 
1 = \int\limits_0^1 dx \, (1-x)^{2 \tau - 3} \, 
\biggl[ 2 f_\tau(x) (\tau-1) - f'_\tau(x) (1-x) \biggr] \,. 
\en 
Here and in the following the superscript $(')$ means derivative 
with respect to variable $x$. 
Using a general definition for the hadronic PDF $q_\tau(x)$, 
in the form of the integral representation (first moment) over $x$  
\eq 
1 = \int\limits_0^1 dx \, q_\tau(x) 
\en 
we get: 
\eq\label{qtau}
q_\tau(x) = (1-x)^{2 \tau - 3} \,
\biggl[ 2 f_\tau(x) (\tau-1) - f'_\tau(x) (1-x) \biggr] 
= \biggl[ - f_\tau(x) (1-x)^{2 \tau - 2} \biggr]' 
\,. 
\en 
We require that the hadronic PDF $q_\tau(x)$ must have the correct 
scaling at large $x$ and this behavior is governed by the 
profile function $f_\tau(x)$. 

\subsection{Pion PDF}

Now let us consider applications. First we look at the pion PDF
at leading twist $\tau = 2$: 
\eq\label{qpi_ADS} 
q_\pi(x) = (1-x)^2 \, \biggl[  \frac{2 f_\pi(x)}{1-x} - f'_\pi(x) \biggr] 
= [-f_\pi(x) (1-x)^2]' 
\en  
Following the pQCD prediction presented in Ref.~\cite{Aicher:2010cb}, 
we consider the parametrization for the pion PDF at the initial scale  
$\mu_0 = 0.63$ GeV as: 
\eq\label{qpi_QCD} 
q_\pi(x,\mu_0) = N_\pi x^{\alpha-1} \, (1-x)^\beta \, (1+\gamma x^\delta) \,, 
\en 
where $N_\pi$ is the normalization constant, 
$\alpha = 0.70$, $\beta = 2.03$, $\gamma = 13.8$, $\delta = 2$. 
Notice that in Ref.~\cite{Gutsche:2014zua} we derived the LF wave function 
which produces this PDF. Now we are on the position to fix the profile 
function $f_\pi(x)$, matching Eqs.~(\ref{qpi_ADS}) and~(\ref{qpi_QCD}). 
Restricting to leading twist, with good accuracy 
we can use an approximate value of the parameter $\beta \simeq 2$ 
in Eq.~(\ref{qpi_QCD}). With this and the boundary condition $f_\pi(0) = 1$ 
we fix $f_\pi(x)$: 
\eq\label{fpi_PDF} 
f_\pi(x) (1-x)^2 = 1 - N_\pi \, x^\alpha 
\biggl[ \frac{1}{\alpha} 
- \frac{2x}{\alpha+1} + \frac{x^2}{\alpha+2}  
+ \gamma x^{\delta} \, 
\biggl( \frac{1}{\alpha+\delta} 
- \frac{2x}{\alpha+\delta+1} + \frac{x^2}{\alpha+\delta+2} \biggr) 
\biggr] \,. 
\en 
It is easy to verify that $f_\pi$ obeys the boundary conditions   
$f_\pi(0) = 1$ and $f_\pi(1) = 0$. At large $x$ it scales 
as $f_\pi(x) \sim (1-x)$, which leads to the correct scaling of the pion PDF: 
$q_\pi(x) \sim (1-x)^2$. We can also write down the relation of function $f_\pi(x)$ 
with $y_\pi(x) \equiv y_2(x)$: 
\eq 
y_\pi(x) = 1 - f_\pi(x) (1-x)^2 \,,
\en 
which for large $x$ due to $f_\pi(x) \sim (1-x)$ simplifies to 
\eq\label{ypi_x_large} 
y_\pi(x) = 1 - (1-x)^3 = x \, (3 - 3 x + x^2) \,. 
\en 

In Ref.~\cite{Gutsche:2012bp} 
we proposed a formalism for the inclusion of high-Fock states in soft-wall 
AdS/QCD. In the case of PDF it is given by the sum: 
\eq
q_\pi(x) = \sum\limits_{\tau = 2, 4, \ldots} \, c_\tau q_\tau(x)\,,
\en 
where $c_\tau$ is the set of mixing coefficients defining the partial 
contributions to the pion PDF, from specific twists $\tau = 2, 4, \ldots$, 
which obey the normalization condition: 
\eq 
1 = \int\limits_0^1 dx \, q_\pi(x) = 
\sum\limits_{\tau = 2, 4, \ldots} \, c_\tau \, 
\int\limits_0^1 dx \, q_\tau(x)  = \sum\limits_{\tau = 2, 4, \ldots} \, c_\tau \,. 
\en

\subsection{Nucleon PDFs}

Next we consider the $u$ and $d$ quark PDFs in the nucleon. The nucleon PDFs and GPDs 
in soft-wall model were calculated for the first time in Ref.~\cite{Vega:2010ns}. 
They were extracted from nucleon electromagnetic form factors using an integral 
presentation for the vector field dual to the electromagnetic current~(\ref{VInt}). 
As we stressed in the Introduction, in previous papers using the soft-wall model, 
the variable of integration in Eq.~(\ref{VInt}) was identified with the
light-cone variable. It led to the results for the PDF and GPDs with much harder 
scaling at large $x \to 1$, i.e. $(1-x)^{\tau-2}$ instead of $(1-x)^{2\tau -3}$. 
To solve this problem, one can identify the variable of integration in Eq.~(\ref{VInt}) 
with arbitrary function of $x$, i.e. with $y_\tau(x)$, and fix $y_\tau(x)$ to guarantee the 
consistency of power scaling of PDFs and GPDs with model-independent results known from QCD. 
One of the solutions for $y_\tau(x)$ consistent with power counting is~\cite{Lyubovitskij:2013ski}:
\eq 
y_\tau(x) = \exp\Big[-\log(1/x) (1-x)^{1/(\tau-2)}\Big]
\en 
leading to the correct large-$x$ scaling of PDFs and GPDs  
\eq 
q_\tau(x) \sim {\cal H}_\tau^\pi(x,Q^2) \sim (1-x)^{2 \tau - 3} \,. 
\en 

As we pointed out before, we follow this novel idea in order to introduce the profile function 
in the normalization condition for the $z$ profiles of the AdS field dual to corresponding 
hadron wave function. Following the pion example considered above, we derive nucleon PDFs 
starting from the normalization conditions, and consistent with model-independent 
counting rules. 
In the nucleon case there are two 
holographic functions dual to its right- $(f^R_\tau)$ and left-chirality 
$(f^L_\tau)$ wave functions~(\ref{fLR_tau}). 
The normalization conditions for the $u$ and $d$ quark wave functions, which are 
equivalent to the normalization conditions for their valence PDFs [$u_v(x)$ 
and $d_v(x)$] read: 

$u$-quark: 
\eq\label{uPDF} 
2 = \int\limits_0^1 dx \, u_v(x) =  
\int\limits_0^\infty dz \, 
\biggl[ 2 \Phi^+(z) \,+\, \eta_u \, 
\partial_z \Big[ z \,  \Phi^-(z) \Big] \biggr]
\en  
$d$-quark: 
\eq\label{dPDF}  
1 = \int\limits_0^1 dx \, d_v(x) =  
\int\limits_0^\infty dz \, 
\biggl[ \Phi^+(z) \,+\, \eta_d \, 
\partial_z \Big[ z \,  \Phi^-(z) \Big] \biggr]
\en  
where 
\eq 
\Phi^\pm = \frac{1}{2} \, \biggl[
\Big(f^R_{\tau}\Big)^2 \,\pm\, \Big(f^L_{\tau}\Big)^2 \biggr] \,, 
\en 
are the combinations of right and left holographic wave functions, 
$\eta_u = 2 \eta_p + \eta_n$ and $\eta_d = 2 \eta_n + \eta_p$  
are the linear combinations of the nucleon couplings with vector field
related to nucleon anomalous magnetic moments $k_N$ and fixed 
as~\cite{Abidin:2009hr,Vega:2010ns}: $\eta_N = k_N \kappa/(2 M_N \sqrt{2})$, 
where $M_N$ is the nucleon mass. 

Notice that the contribution of ``nonminimal'' terms vanish in the normalization 
condition for wave functions and PDFs due gauge invariance, but they contribute 
to the $x$-dependence of PDFs. Moreover, as seen from Eqs.~(\ref{uPDF}) and~(\ref{dPDF}),   
the ``nonminimal contributions'' to the quark PDFs are sufficient to violate the 
symmetry condition $u_v(x)/d_v(x) = 2$, which occurs at $\eta_p = \eta_n = 0$. 

For arbitrary twist the expressions for the 
quark PDFs in the nucleon are given in Appendix~\ref{appendix}.   
For leading twist $\tau=3$, the results for $u_v(x)$ and $d_v(x)$ 
read: 
\eq\label{udv}
u_v(x) &=& \biggl[ 
- f_u(x) (1-x)^4 \, \Big(1 + 2 \eta_u + (1-x)^2 (1 - 4 \eta_u) 
+ 2 \eta_u (1-x)^4 \Big) \biggr]'  \,, \nonumber\\
d_v(x) &=& \biggl[ 
- f_d(x) (1-x)^4 \, \Big(\frac{1}{2} + 2 \eta_d + (1-x)^2 
\Big(\frac{1}{2} - 4 \eta_d\Big) 
+ 2 \eta_d (1-x)^4 \Big) \biggr]'  \,. 
\en 
Both PDFs in Eqs.~(\ref{udv}) scale at large $x$ as $(1-x)^3$, as dictated by 
the counting rules~\cite{Drell:1969km,Blankenbecler:1974tm,Yuan:2003fs}, 
when the $f_u(x)$ and $f_d(x)$ go to constants independent on $x$. 
In other words, the Taylor expansion for $f_q(x)$, $q=u,d$ has the generic form 
\eq 
f_q(x) = \sum\limits_{n} \, c_n (1-x)^n \,, 
\en 
with $\sum\limits_{n} \, c_n = 1$, due to the boundary condition $f_q(0) = 1$. 
Here the sum over $n$ starts from $n=0$. 

World data analysis (see, e.g., Ref.~\cite{Martin:2009iq}) 
supports the $(1-x)^3$ scaling of the $u_v$ PDF, while the extracted $d_v$ 
PDF has softer behavior $(1-x)^5$. Note that other groups give
either similar fits or softer behavior of the $d$ quark PDF, 
such as e.g. $(1-x)^{4.47 \pm 0.55}$~\cite{Alekhin:2017kpj}, or 
introduce into the $d$ quark PDF a nontrivial polynomial depending 
on $\sqrt{x}$~\cite{Hou:2019efy}.  
In our approach we can resolve this  
puzzle. The solution is based on a suppression of $(1-x)^3$ term in $d_v$ 
[see Eq.~(\ref{udv})], which can occur when the following constraint 
on the $\eta_d$ coupling holds:
\eq\label{dv_condition} 
\frac{1}{2} + 2 \eta_d = 0 \,.
\en 
From the latter condition it follows that the dilaton scale parameter $\kappa$ 
is related to the nucleon mass as $\kappa = 0.348 \, M_N = 326$ MeV, 
which is very close to the value 
$\kappa = 350$ MeV used in Refs.~\cite{Abidin:2009hr,Vega:2010ns}. 
Adopting the condition~(\ref{dv_condition}) and restricting to the leading order in 
the $(1-x)$ expansion, we get the following expressions for the quark PDFs 
in the nucleon: 
\eq\label{udv_fin} 
u_v(x) = \Big[-f_u(x) (1-x)^4\Big]' 
\,, \quad 
d_v(x) = \Big[-f_d(x) (1-x)^6\Big]' \,. 
\en 
Now we fix the $u$ and $d$ profile functions $f_u(x)$ and $f_d(x)$, using 
predictions for the valence PDFs $u_v(x)$ and $d_v(x)$ extracted from 
world data analysis. As an example, we use the results of 
the MSTW 2008 LO global analysis~\cite{Martin:2009iq}: 
\eq
u_v(x,\mu_0) &=& A_u \, x^{\alpha_u-1} \, (1-x)^{\beta_u} \, 
(1 + \epsilon_u \sqrt{x} + \gamma_u x)\,, \\
d_v(x,\mu_0) &=& A_d \, x^{\alpha_d-1} \, (1-x)^{\beta_d} \, 
(1 + \epsilon_d \sqrt{x} + \gamma_d x)\,, 
\en
where $\mu_0 = 1$ GeV is the initial scale. 
The normalization constants $A_q$ 
and the constants $\alpha_q$, $\beta_q$, $\epsilon_q$, $\gamma_q$ 
were fixed as 
\eq 
& &A_u = 1.4335\,, \quad A_d = 5.0903\,, 
\nonumber\\
& &\alpha_u = 0.45232\,, \quad \alpha_d = 0.71978\,, 
\nonumber\\
& &\beta_u = 3.0409 \simeq 3\,, \quad \beta_d = 5.1244 \simeq 5\,, 
\\
& &\epsilon_u = -2.3737\,, \quad \epsilon_d = -4.3654\,, 
\nonumber\\
& &\gamma_u = 8.9924\,, \quad \gamma_d = 7.4730 \,.
\nonumber
\en 
Solving the differential equations~(\ref{udv_fin}) 
with the boundary condition $f_q(0) = 1$ and 
using $\beta_u = 2$, $\beta_d =5$ we get: 
\eq 
f_u(x) \, (1-x)^4 &=& 1 - A_u x^{\delta_u} \, 
\Big[                 B_u(x,0) 
+ \epsilon_u \sqrt{x} B_u(x,1/2)
+        \epsilon_u x B_u(x,1) \Big] \,, \\
f_d(x) \, (1-x)^6 &=& 1 - A_d x^{\delta_d} \, 
\Big[                 B_d(x,0) 
+ \epsilon_d \sqrt{x} D_d(x,1/2)
+        \epsilon_d x B_d(x,1) \Big] \,, 
\en 
where 
\eq
B_u(x,n) &=&  \sum\limits_{k=0}^3 \, \frac{C_3^k \, (-x)^k}{\delta_u+n+k} =
     \frac{1}{\delta_u+n} 
     -    \frac{3 x}{\delta_u+n+1} 
     +  \frac{3 x^2}{\delta_u+n+2} 
     -    \frac{x^3}{\delta_u+n+3}   \,, \\
B_d(x,n) &=&  \sum\limits_{k=0}^5 \, \frac{C_5^k \, (-x)^k}{\delta_d+n+k} =
     \frac{1}{\delta_d+n} 
     -    \frac{5 x}{\delta_d+n+1} 
     + \frac{10 x^2}{\delta_d+n+2} 
     - \frac{10 x^3}{\delta_d+n+3}  
     +  \frac{5 x^4}{\delta_d+n+4}  
     -    \frac{x^5}{\delta_d+n+5}  \,. 
\en
Here $C_m^k = \frac{m!}{k! (m-k)!}$ are the binomial coefficients. 
As in the pion case, we derive the relations between sets of 
nucleon functions $y_q(x)$ and $f_q(x)$: 
\eq\label{yf_ud} 
y_u(x) = 1 - \sqrt{f_u(x)} \, (1-x)^2 \,, \qquad 
y_d(x) = 1 - \sqrt{f_d(x)} \, (1-x)^3 \,. 
\en 
For large $x \to 1$ the expressions for $f_q(x)$, and therefore  
the relations~(\ref{yf_ud}), are simplified: 
\eq\label{fy_nucleon}  
f_u(x) &=& f_d(x) = 1 \,, \nonumber\\
y_u(x) &=& 1 - (1-x)^2 = x (2 - x) \,, \\
y_d(x) &=& 1 - (1-x)^3 = x (3 - 3x + x^2) \,. \nonumber 
\en 
It is clear that in this limit the quark PDFs in the nucleon obey the correct 
large $x$ scaling: 
\eq 
u_v(x) = 8 \, (1-x)^3\,, \qquad d_v(x) = 6 \, (1-x)^5 \,.
\en  
Note that we use the MSTW 2008 LO global analysis as an example 
of application of our framework. We can choose any other 
and match the profile functions $f_q$ accordingly. The universality 
of our approach is that the profile functions $f_q$ appear 
in other parton distributions like TMDs and GPDs. Therefore, 
as soon as the profile functions $f_q$ are fixed from PDFs, 
one can have predictions for the other parton densities. 

Now we turn to a discussion of the magnetization PDFs in 
nucleons ${\cal E}_v^u(x)$ and 
${\cal E}_v^d(x)$. The idea of their derivation is similar 
to the case of the charged 
PDFs $u_v(x)$ and $d_v(x)$. We start with expressions for the contribution to 
the anomalous magnetic moments $k^q$ of $u$ and $d$ quarks in soft-wall AdS/QCD 
model~\cite{Vega:2010ns,Abidin:2009hr}, given as integrals over 
left- and right-chirality 
nucleon wave functions with specific twist $\tau$~(\ref{fLR_tau}): 
\eq\label{normmag_phi}
k^q_\tau = 2 M_N \eta_q \, \int\limits_0^\infty dz \, z \, 
\phi_\tau^L(z) \, \phi_\tau^R(z) 
         = \frac{2 M_N}{\kappa} \, \eta_q \, \sqrt{\tau-1}\,. 
\en 
Next we use the integral representation for unity~(\ref{int_unity}) and after integration 
of the holographic variable $z$, we get the magnetization PDFs in the nucleon for 
leading twist $\tau = 3$ [expressions for arbitrary twist can be found in 
Appendix~\ref{appendix}]: 
\eq 
{\cal E}_v^q(x) = k^q \, \Big[ - f_q(x) \, (1-x)^{6} \Big]'  \,. 
\en 
In principle, the $f_q(x)$ profile functions can be different in charged and 
magnetization PDFs. In the case when they are the same we derive the following relation: 
\eq 
\frac{{\cal E}_v^d(x)}{d_v(x)} = 4 \eta_d \, \frac{M_N}{\kappa} \,. 
\en 

\section{TMD}   

TMD can arise in soft-wall AdS/QCD by analogy with PDF, 
using the generalized integral representation for unity,  
including integration over the longitudinal $x$ 
and transverse $\bfk$ variables:  
\eq\label{int_unity2} 
1 &=& - e^{\kappa^2 z^2} \, 
\int\limits_0^1 d\biggl[f_\tau(x) 
e^{-\kappa^2 z^2/(1-x)^2} \biggr] \, 
\int d^2\bfk \, \frac{D_\tau(x)}{\pi \kappa^2} \, 
e^{-\bfk^2 D_\tau(x)/\kappa^2} \nonumber\\ 
&=& \frac{e^{\kappa^2 z^2}}{\pi \kappa^2} \,   
\int\limits_0^1 dx \, \int d^2\bfk \, \biggl[ 
\frac{2f_\tau(x) \, \kappa^2 z^2}{(1-x)^3} \, - f'_\tau(x) \biggr] 
\, D_\tau(x) \, 
e^{-\kappa^2 z^2/(1-x)^2} \, 
e^{-\bfk^2 D_\tau(x)/\kappa^2}  \,, 
\en  
where $D_\tau(x)$ is the longitudinal factor derived in Ref.~\cite{Gutsche:2016gcd}, 
which was fixed from data on the nucleon electromagnetic form factors. 
The purpose of the function $D_\tau(x)$ is to include a running scale in TMD, 
i.e. scale parameter, which accompanies the $\bfk$ dependence in TMDs. In our case 
the running scale parameter is $\Lambda_\tau(x) = \kappa/\sqrt{D_\tau(x)}$. 
As was shown in Ref.~\cite{Gutsche:2016gcd}, 
the appearance of the $\kappa$ or $M_N$ in $\Lambda_\tau(x)$ 
is for convenience, because any different choice can be compensated by rescaling the 
function $D_\tau(x)$. Such choice of $\Lambda_\tau(x)$ is a generalization of the
Gaussian ansatz for TMD with constant scale $\Lambda^2 = \langle \bfk^2 \rangle$
in the exponential, proposed by Turin group~\cite{Anselmino:2002pd}:
\eq\label{TMD_ansatz}
F(x,\bfk) = F(x) \, e^{-\bfk^2/\langle \bfk^2 \rangle} \,.
\en
This Gaussian ansatz~(\ref{TMD_ansatz}) is simple and very useful 
in practical calculations and analysis of data. However, it is known (see e.g.,
Ref.~\cite{Bacchetta:2019tcu}), that it presents difficulties in the description
of data on DY processes in some kinematical regions (e.g. at $Q_\perp \le Q$).
Therefore, the ansatz for the TMD~(\ref{TMD_ansatz})  
can be crucially checked. In this vein, one can mention results
of AdS/QCD and light-front quark models motivated by AdS/QCD
(see Refs.~\cite{Gutsche:2013zia,Gutsche:2016lrz,Gutsche:2016gcd})
where it was shown that the hadronic light-front wave functions,
PDFs, and TMDs contain scale parameter depending on the light-cone variable $x$,
i.e. they can be considered as $x$-dependent scale quantities  
It was found in Refs.~\cite{Gutsche:2013zia,Gutsche:2016lrz,Gutsche:2016gcd}  
that $x$-dependent scale is crucial for a successful description of data 
on electromagnetic 
form factors of nucleons and electroexcitation of nucleon resonances. Also we can see 
below that our result for the unpolarazed quark TMD in nucleon 
will contain two terms multiplied with a Gaussian: constant term and term proportional 
to $\bfk^2$. It is consistent with the form of TMD used 
by the Pavia group~\cite{Bacchetta:2017gcc}. 
In the next section we will show that function $D_\tau(x)$ can be fixed 
from expression for the electromagnetic form factor and related 
to functions $f_\tau(x)$ and $y_\tau(x)$. 

Using the same calculation technique as for the case of PDFs, we insert the integral 
representation~(\ref{int_unity2}) into the normalization condition for the 
holographic wave function~(\ref{norm_phi}) and integrate over the $z$ variable. 
After that we arrive at the normalization condition for the TMD $F_\tau(x,\bfk)$, 
from which the latter can be extracted and expressed through PDF as: 
\eq\label{phiR_kp}
1 = \int\limits_0^1 dx \, \int d^2\bfk \, F_\tau(x,\bfk)\,, 
\qquad 
F_\tau(x,\bfk) =  q_\tau(x) \, \frac{D_\tau(x)}{\pi \kappa^2} 
\, e^{-\bfk^2 D_\tau(x)/\kappa^2}  \,.   
\en 
Also it is important to stress that from the results for generic PDFs 
and TMDs derived in 
present paper one can set up LF quark model in analogy with our previous 
papers~\cite{Gutsche:2013zia,Gutsche:2016gcd}. In particular, 
the LF wave function for generic hadron with twist $\tau$ reads: 
\eq\label{LFWF_tau}
\psi(x,\bfk) = \frac{4\pi}{\kappa} \, \sqrt{q_\tau(x) \, D_\tau(x)}  \, 
\exp\biggl[- \frac{\bfk^2}{2 \kappa^2} \, D_\tau(x) \biggr] \,. 
\en 
Note that generic TMD and PDF 
are expressed in term of LF wave function~(\ref{LFWF_tau}) as: 
\eq 
F_\tau(x,\bfk) = \frac{1}{16 \pi^3} \, |\psi(x,\bfk)|^2\,, \qquad 
q_\tau(x) = \int \frac{d^2\bfk}{16 \pi^3}\, |\psi(x,\bfk)|^2 
          = \int d^2\bfk \, F_\tau(x,\bfk) \,. 
\en 
Now lets consider as example the result for the unpolarized quark TMD in nucleon 
$f_1^{q_v}(x,\bfk)$. 
As in case of PDF it is contributed by two wave functions $\phi^R(z)$ and $\phi^L(z)$~(\ref{fLR_tau}) 
corresponding to the leading and subleading twist or having orbital moment $L=0$ and 
$L=1$. The $\phi^R(z)$ function generates the contribution to TMD $f_{1,R}^{q_v}(x,\bfk)$ 
fixed from condition similar to Eq.~(\ref{phiR_kp}), while the $\phi^L(z)$ gives the contribution 
$f_{1,L}^{q_v}(x,\bfk)$ proportional to $\bfk^2$: 
\eq 
f_1^{q_v}(x,\bfk) = f_{1,R}^{q_v}(x,\bfk) + f_{1,L}^{q_v}(x,\bfk)
\,, 
\en     
where 
\eq\label{phiL_kp}
f_{1,R}^{q_v}(x,\bfk) = q_v^+(x) 
\, \frac{D_q(x)}{2 \pi \kappa^2} 
\, e^{-\bfk^2 D_q(x)/\kappa^2}  \,,  \quad 
f_{1,L}^{q_v}(x,\bfk)  = q_v^-(x) 
\, \frac{\bfk^2 D_q^2(x)}{2 \pi \kappa^4} 
\, e^{-\bfk^2 D_q(x)/\kappa^2}  \,, 
\en 
Here $q_v^\pm(x) = q_v(x) \pm \delta q_v(x)$, 
$q_v(x)$ and $\delta q_v(x)$ are the helicity-independent 
and helicity-dependent valence quark parton distributions. 
As we mentioned before, the form of our expression for TMD 
\eq 
f_1^{q_v}(x,\bfk) &=& 
\biggl[q_v^+(x) + q_v^-(x) \, 
\frac{\bfk^2 \, D_q(x)}{\kappa^2}\biggr] \, 
\frac{D_q(x)}{2 \pi \kappa^2} \, 
e^{-\bfk^2 D_q(x)/\kappa^2}   
\en 
is very similar 
to the parametrization used by Pavia group~\cite{Bacchetta:2017gcc}:  
\eq
f_1^a(x,\bfk) = \frac{1}{\pi g_{1a}} \, 
\frac{1 + \lambda \bfk^2}{1 + \lambda g_{1a}} 
\, e^{-\bfk^2/g_{1a}} \,. 
\en 
Using expressions for nucleon PDFs and TMDs one can 
set up the LF wave functions for the nucleon following 
Refs.~\cite{Gutsche:2013zia,Gutsche:2016gcd}: 
\eq\label{LFWF_in}
\psi_{\pm q}^\pm(x,\bfk) = \varphi_q^{(1)}(x,\bfk)
\,, \qquad 
\psi_{\mp q}^\pm(x,\bfk) = \mp \frac{k^1 \pm ik^2}{M_N}
\, \varphi_q^{(2)}(x,\bfk) \,, 
\en
where 
\eq
\varphi_q^{(1)}(x,\bfk) &=& \frac{2 \pi \sqrt{2}}{\kappa} \,
\sqrt{q_v^+(x) \, D_q(x)}  \, 
\exp\biggl[- \frac{\bfk^2}{2 \kappa^2} \, D_q(x) \biggr] \,, 
\nonumber\\
\frac{1}{M_N} \, 
\varphi_q^{(2)}(x,\bfk) &=& \frac{2 \pi c_q \sqrt{2}}{\kappa^2} \, 
\sqrt{q_v^-(x)} \, D_q(x) \, 
\exp\biggl[- \frac{\bfk^2}{2 \kappa^2} \, D_q(x) \biggr] \,. 
\en
Here $c_u=1$, $c_d=-1$, $\psi_{\lambda_q q}^{\lambda_N}(x,\bfk)$
are the LFWFs at the initial scale $\mu_0$ with specific helicities for
the nucleon $\lambda_N  = \pm$ and for the struck quark $\lambda_q = \pm $,
where plus and minus correspond to $+\frac{1}{2}$ and $-\frac{1}{2}$,
respectively. Note, in terms LF wave functions~(\ref{LFWF_in}) the unpolarized 
quark TMD in nucleon reads~\cite{Bacchetta:2008af}:
\eq
f_1^{q_v}(x,\bfk) &=& \frac{1}{16 \pi^3} \, \biggl[
|\psi_{+q}^+(x,\bfk)|^2
+ |\psi_{-q}^+(x,\bfk)|^2  \biggr] =
\frac{1}{16 \pi^3} \, \biggl[
\Big(\varphi_q^{(1)}(x,\bfk)\Big)^2
+ \frac{\bfk^2}{M_N^2}\Big(\varphi_q^{(2)}(x,\bfk)\Big)^2 \biggr]
\,.
\en
Note $q_v^\pm(x)$ and ${\cal E}_v^q(x)$ PDFs are related as~\cite{Gutsche:2016gcd}:  
\eq
{\cal E}_v^q(x) = c_q \, \sqrt{q_v^+(x) \, q_v^-(x) \, D_q(x)} \, (1-x) \,.
\en 
The full set of the valence $T$-even TMDs generated by LF wave functions derived above 
is listed in Appendix~\ref{app_TMD}. 

\section{GPD} 

As we mentioned before, the nucleon GPDs were calculated for the first time 
in soft-wall AdS/QCD in Ref.~\cite{Vega:2010ns}. These quantities were expressed 
in terms of {\it generalized light-cone variable} $y_\tau(x)$, which has direct 
relation to the profile function $f_\tau(x)$. Function $f_\tau(x)$ is more convenient for 
displaying power behavior of hadronic parton distributions (PDFs, TMDs, and GPDs). 
In particular, for arbitrary twist $\tau$, 
a generic GPD in hadron reads~\cite{Gutsche:2013zia}: 
\eq 
{\cal H}_\tau(y_\tau(x),Q^2) = (\tau - 1) \, 
(1-y_\tau(x))^{\tau-2} \, \Big[y_\tau(x)\Big]^a\,, \quad a = \frac{Q^2}{4 \kappa^2} \,. 
\en 
It can be written in more convenient form  in terms of PDF: 
\eq 
{\cal H}_\tau(x,Q^2) = q_\tau(x) \, \Big[y_\tau(x)\Big]^a = 
q_\tau(x) \, \exp\Big(- a \log\Big[1/y_\tau(x)\Big]\Big) \,, 
\en  
where the PDF $q_\tau(x)$ and light-cone function $y_\tau(x)$ are expressed through profile 
function $f_\tau(x)$ according to Eqs.~(\ref{qtau}) and~(\ref{yf_relation}). 

Next we constrain function $D_\tau(x)$ and relate it to functions $y_\tau(x)$ and 
$f_\tau(x)$ matching the expression for the hadronic form factors in two 
approaches --- soft-wall AdS/QCD and LF QCD. The LF QCD result for the hadron form factor 
is given by the DYW formula~\cite{Drell:1969km}
\eq\label{DYW_f}
F_\tau(Q^2) = \int\limits_0^1 dx \, \int\frac{d^2\bfk}{16\pi^3} \,
\psi^\dagger_\tau(x,\bfk') \, \psi_\tau(x,\bfk) \,,
\en
where $\psi(x,\bfk) \equiv  \psi(x,\bfk; \mu_0)$ 
is wave function  derived in Eq.~(\ref{LFWF_tau}), 
$\bfk' = \bfk + (1-x) \bfq$, and $Q^2 = \bfq^2$. 

We get:  
\eq 
F_\tau(Q^2) = \int\limits_0^1 dx \, q_\tau(x) \, \exp\Big[- a \log[1/y_\tau(x)]\Big] 
            = \int\limits_0^1 dx \, q_\tau(x) \, \exp\Big[- a D_\tau(x) (1-x)^2\Big] 
\en 
or 
\eq 
D_\tau(x) = \frac{1}{(1-x)^2} \, \log[1/y_\tau(x)]= 
\frac{1}{(1-x)^2} \, 
\log\biggl[1 - \Big(f_\tau(x)\Big)^{\frac{1}{\tau-1}} \, (1-x)^2 \biggr]^{-1} \,. 
\en 
For large $x$ function $D_\tau(x)$ behaves as 
\eq 
D_\tau(x) = \Big(f_\tau(x)\Big)^{\frac{1}{\tau-1}} \,, 
\en 
where $f_\pi(x) = 1-x$, $f_u(x) = f_d(x) = 1$ and therefore 
$D_\pi(x) = 1-x$, $D_u(x) = D_d(x) = 1$. It leads to the following 
scaling of the TMDs at large $x$: 
\eq
f_1^\pi(x,\bfk) = q_\pi(x) \, (1-x)  
\, \frac{e^{-\bfk^2 (1-x)/\kappa^2}}{\pi \kappa^2} 
\en 
for pion, 
\eq 
f_1^{q_v}(x,\bfk) = \biggl[
q_v^+(x) + q_v^-(x) \, \frac{\bfk^2}{\kappa^2}
\biggr] \, \frac{e^{-\bfk^2/\kappa^2}}{2 \pi \kappa^2} 
\en 
for nucleon. 
 
Now we consider specific cases for GPDs. In the pion case we have $\tau = 2$ and 
$y_\pi(x) = 1 - f_\pi(x) \, (1-x)^2$, where the pion profile function $f_\pi(x)$
is fixed from pion PDF by Eq.~(\ref{fpi_PDF}). The pion PDF $q_\pi(x)$ is fixed from 
data. Therefore, we give the pion GPD prediction at the initial scale $\mu_0 = 1$ GeV  
in terms of the pion PDF, or more precisely in terms of constants parametrizing PDF ($N_\pi$, 
$\alpha$, $\beta$, $\gamma$, $\delta$) fixed in Ref.~\cite{Aicher:2010cb}.   
At large $x$ the profile functions $f_\pi(x) \to (1-x)$ and 
$y_\pi(x) \to 1$ [see Eq.~(\ref{ypi_x_large})], 
and the scaling of our result for the pion GPD $(1-x)^2$ is consistent with the pQCD prediction 
~\cite{Yuan:2003fs}: it coincides with the leading-order result for the
pion PDF and is independent on $Q^2$: 
\eq 
{\cal H}_\pi(x,Q^2) = q_\pi(x) = 3 \, (1-x)^2 \,. 
\en 
In the nucleon case we have $\tau = 3$, 
$y_u(x) = 1 - \sqrt{f_u(x)} \, (1-x)^2$, and 
$y_d(x) = 1 - \sqrt{f_d(x)} \, (1-x)^3$. The quark profile functions  
$f_u(x)$ and $f_d(x)$ are fixed from the corresponding nucleon PDFs 
extracted from global data analysis at the 
initial scale $\mu_0 = 1$ GeV~\cite{Martin:2009iq}. 
The four (charged and magnetization) nucleon GPDs at the initial 
scale $\mu_0 = 1$ GeV  are defined as: 
\eq 
{\cal H}^q_v(x,Q^2) = q_v(x) \, \Big[y_q(x)\Big]^a\,, \qquad 
{\cal E}^q_v(x,Q^2) = {\cal E}_v^q(x) \, \Big[y_q(x)\Big]^a \,. 
\en 
Finally we consider the limit of large $x$. In this case 
the profile functions $f_q(x)$ and functions $y_q(x)$ approach 1: 
$f_u(x) = f_d(x) = 1$ and $y_u(x) = y_d(x) = 1$ [see Eq.~(\ref{fy_nucleon})].  
The scaling of the nucleon charge and magnetization GPDs are also (as in case 
of pion) consistent with the pQCD predictions~\cite{Yuan:2003fs}: 
\eq 
{\cal H}^u_v(x,Q^2) = u_v(x) = 8 \, (1-x)^3\,, \qquad 
{\cal E}^q_v(x,Q^2) = {\cal E}^q_v(x) = 6 \, {\cal E}^q_v \, (1-x)^5\,.
\en 
In the case of the $d$ quark charge GPD ${\cal H}^d_v(x,Q^2)$ we have two possibilities 
at large $x$. In general it scales as $(1-x)^3$ in agreement with pQCD~\cite{Yuan:2003fs}. 
On the other hand, if we suppress the leading-order term $(1-x)^3$ in the $d$ quark PDF 
using the constraint~(\ref{dv_condition}), then $d_v(x)$ 
has softer $(1-x)^5$ behavior consistent with result of 
world data analysis~\cite{Martin:2009iq}. In this vein,  we also get $(1-x)^5$ scaling of 
the ${\cal H}^d_v(x,Q^2)$. Note that the large $x$ scaling of the pion and nucleon GPDs 
is governed by corresponding PDFs.  

\section{Summary}

In the present paper we have explicitly demonstrated how to correctly 
define the hadronic parton distributions (PDFs, TMDs, and GPDs) in the
soft-wall AdS/QCD approach based on the use of quadratic
dilaton. The large $x$ behavior of PDFs and GPDs is consistent with 
model-independent counting rules. For the first time, we derive results for the 
large $x$ behavior of TMDs. Our predictions for the $T$-even TMDs of nucleon 
are listed in Appendix~\ref{app_TMD}. All parton distributions are defined 
in terms of profile functions $f_\tau(x)$  depending on the
light-cone coordinate. The functions $f_\tau(x)$ are related to the PDFs and 
obey the boundary condition $f_\tau(0) = 1$. 
We also proposed a solution to the puzzle related with a softer large $x$ 
behavior of the valence $d$ quark PDF in nucleon in comparison with 
the one of the $u$ quark. It can be obtained due to the vanishing of the 
leading-order term $(1-x)^3$ when nonminimal couplings of the nucleons 
with the electromagnetic field obey the condition~(\ref{dv_condition}). 
Profile functions are fixed from data analysis on PDFs and can then be tested 
in the phenomenology of TMDs and GPDs.  

\appendix
\section{Useful analytical results for parton densities}
\label{appendix} 

For arbitrary twist the expressions for the 
quark PDFs in nucleon read:
\eq
u_v(x) &=& \biggl[ 
- f_u(x) (1-x)^{2 (\tau-1)} \, \Big(1 + \eta_u (\tau-1) + (1-x)^2 (1 - 2 \eta_u (\tau-1)) 
+ \eta_u (\tau-1) (1-x)^4 \Big) \biggr]'  \,, \\
d_v(x) &=& \biggl[ 
- f_d(x) (1-x)^{2 (\tau-1)} \, \Big(\frac{1}{2} + \eta_d (\tau - 1) + (1-x)^2 
\Big(\frac{1}{2} - 2 \eta_d (\tau-1) \Big) 
+ \eta_d (\tau-1) (1-x)^4 \Big) \biggr]'  \,. 
\en 

In the $\tau = 3$ case and using the additional constraint $2 \eta_d = - 1/2$ 
(it means that we get $2 \eta_u = 3 \eta_p - 1/4$),  
we can suppress the leading $(1-x)^3$ term 
in $d_v(x)$. Therefore, $d_v(x)$ dominates by the next-to-leading term $(1-x)^5$.  
Taking all these arguments into account we arrive at:  
\eq 
u_v(x) &=& \biggl[ 
- \frac{3}{4} f_u(x) (1-x)^4 \, \Big(1 + 4\eta_p + 2 (1-x)^2 (1 - 4 \eta_p) 
- \frac{1}{3} (1-x)^4 (1 - 12 \eta_p) \Big) \biggr]'  \,, \\
d_v(x) &=& \biggl[
- \frac{3}{2} f_d(x) (1-x)^6 \, \biggl(1 -  \frac{(1-x)^2}{3}\biggr) 
\biggr]' \,. 
\en 
Restricting for simplicity to the leading order in $(1-x)$ expansion of $u_v(x)$ 
and $d_v(x)$ we finally get 
\eq 
u_v(x) = \Big[-f_u(x) (1-x)^4\Big]' 
\,, \qquad 
d_v(x) = \Big[-f_d(x) (1-x)^6\Big]' \,. 
\en 

Magnetization quark PDFs in nucleon for arbitrary twist are given by 
\eq 
{\cal E}_v^q(x) = k^q \, \Big[ - f_q(x) \, (1-x)^{2 \tau} \Big]'\,, 
\qquad 
k^q = \frac{2 M_N}{\kappa} \, \eta_q \, \sqrt{\tau-1} \,.
\en

\section{$T$-even TMDs of nucleon}
\label{app_TMD} 

Here we list the $T$-even TMDs of nucleon using derived LF
decomposition discussed 
in~\cite{Bacchetta:2008af} and~\cite{Gutsche:2016gcd} and  
wave functions derived in Eq.~(\ref{LFWF_in}): 
\eq
& &f_1^{q_v}(x,\bfk) \equiv h_{1T}^{q_v}(x,\bfk) = 
\frac{1}{16 \pi^3} \, \biggl[
\Big(\varphi_q^{(1)}(x,\bfk)\Big)^2
+ \frac{\bfk^2}{M_N^2}\Big(\varphi_q^{(2)}(x,\bfk)\Big)^2 \biggr]
\,, \nonumber\\
& &g_{1L}^{q_v}(x,\bfk) =
\frac{1}{16 \pi^3} \, \biggl[
\Big(\varphi_q^{(1)}(x,\bfk)\Big)^2
- \frac{\bfk^2}{M_N^2}\Big(\varphi_q^{(2)}(x,\bfk)\Big)^2 \biggr]
\,,
\nonumber\\
& &g_{1T}^{q_v}(x,\bfk) \equiv - h_{1L}^{\perp q_v}(x,\bfk) 
=
\frac{1}{8 \pi^3} \, \varphi_q^{(1)}(x,\bfk)
                  \, \varphi_q^{(2)}(x,\bfk) \,,
\\
& &h_1^{q_v}(x,\bfk) \equiv h_{1T}^{q_v}(x,\bfk)
+ \frac{\bfk^2}{2M_N^2} h_{1T}^{\perp q_v}(x,\bfk) 
=\frac{1}{16 \pi^3} \, \Big(\varphi_q^{(1)}(x,\bfk)\Big)^2 \,,
\nonumber\\
& &\frac{\bfk^2}{2M_N^2} h_{1T}^{\perp q_v}(x,\bfk) =
\frac{1}{2} \Big[g_{1L}^{q_v}(x,\bfk) - f_1^{q_v}(x,\bfk)\Big] 
=  g_{1L}^{q_v}(x,\bfk) - h_1^{q_v}(x,\bfk) 
= - \frac{\bfk^2}{16 \pi^3 M_N^2} \,
\Big(\varphi_q^{(2)}(x,\bfk)\Big)^2 \,. \nonumber 
\en
Using our expressions of the LF wave functions we express TMDs
through the PDFs 
\eq
f_1^{q_v}(x,\bfk) &\equiv& h_{1T}^{q_v}(x,\bfk) 
={\cal F}_1(x,\bfk) + {\cal F}_2(x,\bfk)\,, \nonumber\\
g_{1L}^{q_v}(x,\bfk) &=& {\cal F}_1(x,\bfk) - {\cal F}_2(x,\bfk)\,,
\nonumber\\
g_{1T}^{q_v}(x,\bfk) &\equiv& - h_{1L}^{\perp q_v}(x,\bfk) =
{\cal F}_3(x,\bfk) \,, \\
h_1^{q_v}(x,\bfk) &=& {\cal F}_1(x,\bfk)\,,
\nonumber\\
\frac{\bfk^2}{2M_N^2} h_{1T}^{\perp q_v}(x,\bfk) &=&
- {\cal F}_2(x,\bfk)\,, \nonumber
\en
where
\eq
{\cal F}_1(x,\bfk) &=& 
q_v^+(x) \, \frac{D_q(x)}{2\pi\kappa^2} \, 
e^{- \frac{\bfk^2}{\kappa^2} D_q(x)}\,, \nonumber\\
{\cal F}_2(x,\bfk) &=& 
q_v^-(x) \, \frac{\bfk^2 D_q^2(x)}{2\pi\kappa^4} \,  
e^{- \frac{\bfk^2}{\kappa^2} D_q(x)}\,, \nonumber\\
{\cal F}_3(x,\bfk) &=& c_q \, 
\sqrt{\frac{4\kappa^2}{\bfk^2} \,
{\cal F}_1(x,\bfk) \, {\cal F}_2(x,\bfk) } 
= \sqrt{q_v^+(x) \, q_v^-(x)} \, 
\frac{c_q \, D_q^{3/2}(x)}{\pi \kappa^2} \, 
e^{- \frac{\bfk^2}{\kappa^2} \, D_q(x)}\,.
\en
Performing the $\bfk$-integration over the TMDs with
\eq
{\rm TMD}(x) = \int d^2\bfk \, {\rm TMD}(x,\bfk) \,, \qquad 
\overline{{\rm TMD}}(x) = \int d^2\bfk \, \frac{\bfk^2}{2M_N^2} \,
{\rm TMD}(x,\bfk)
\en
gives the identities
\eq
& &f_1^{q_v}(x) \equiv h_{1T}^{q_v}(x) \, = \, q_v(x)\,,
\qquad
g_{1L}^{q_v}(x) = \delta q_v(x)\,,
\qquad 
g_{1T}^{q_v}(x) \equiv - h_{1L}^{\perp q_v}(x) =
\frac{{\cal E}^q(x)}{1-x} \,,
\nonumber\\
& &h_1^{q_v}(x) = \frac{q_v(x) + \delta q_v(x)}{2}\,, 
\qquad 
\overline{h_{1T}^{\perp q_v}}(x) = 
- \frac{q_v(x) - \delta q_v(x)}{2}\,.
\en
The integration over $x$ leads to the normalization conditions
\eq
\int\limits_0^1 dx f_1^{q_v}(x) = 
\int\limits_0^1 dx h_{1T}^{q_v}(x) = n_q \,,
\quad 
\int\limits_0^1 dx g_{1L}^{q_v}(x) = g_A^q \,,
\quad 
\int\limits_0^1 dx h_1^{q_v}(x) = g_T^q \,,
\en
where $n_q$ is the number of $u$ or $d$ valence quarks in the proton, 
$g_A^q$ is the axial charge of a quark with flavor $q=u$ or $d$, and 
$g_T^q$ is the tensor charge. Our TMDs satisfy all relations and 
inequalities found before in theoretical approaches 
(see detailed discussion in~Ref.~\cite{Gutsche:2016gcd}).

\begin{acknowledgments}

This work was funded 
by ``Verbundprojekt 05P2018 - Ausbau von ALICE 
am LHC: Jets und partonische Struktur von Kernen'' 
(F\"orderkennzeichen: 05P18VTCA1), 
by ``Verbundprojekt 05A2017 - CRESST-XENON: 
Direkte Suche nach Dunkler 
Materie mit XENON1T/nT und CRESST-III. Teilprojekt 1''
(F\"orderkennzeichen 05A17VTA)'', by CONICYT (Chile) under
Grants No. 7912010025, No. 1180232 and ANID PIA/APOYO AFB180002 
and by FONDECYT (Chile) under Grant No. 1191103. 

\end{acknowledgments}

\end{document}